# Spontaneous doping of the basal plane of MoS$_2$ single-layers through oxygen substitution under ambient conditions

*János Pető[1], Tamás Ollár[2], Péter Vancsó[1,3], Zakhar I. Popov[4], Gábor Zsolt Magda[1], Gergely Dobrik[1], Chanyong Hwang[5], Pavel B. Sorokin[4], and Levente Tapasztó[1*]*

1. Hungarian Academy of Sciences, Centre for Energy Research, Institute of Technical Physics and Materials Science, 2DNanoelectronics Lendület Research Group 1121 Budapest, Hungary
2. Hungarian Academy of Sciences, Centre for Energy Research, Institute for Energy Security and Environmental Safety, Surface Chemistry and Catalysis Department 1121 Budapest, Hungary
3. University of Namur, Department of Physics, 61 rue de Bruxelles, 5000 Namur, Belgium
4. National University of Science and Technology MISiS, 119049 Moscow, Russia
5. Korea Research Institute for Standards and Science, Daejeon 305340, South Korea

* *tapaszto@mfa.kfki.hu*

**The chemical inertness of the defect-free basal plane confers environmental stability to MoS$_2$ single-layers, but it also limits their chemical versatility and catalytic activity. The stability of the pristine MoS$_2$ basal plane against oxidation under ambient conditions is a widely accepted assumption in the interpretation of various studies and applications. However, single-atom level structural investigations reported here reveal that oxygen atoms spontaneously incorporate into the basal plane of MoS$_2$ single layers during ambient exposure. Our scanning tunneling microscopy investigations reveal a slow oxygen substitution reaction, upon which individual sulfur atoms are one by one replaced by oxygen, giving rise to solid solution type 2D MoS$_{2-x}$O$_x$ crystals. O substitution sites present all over the basal plane act as single-atomic active reaction centers, substantially increasing the catalytic activity of the entire MoS$_2$ basal plane for the electrochemical H$_2$ evolution reaction.**

Layered materials display thickness-dependent properties when approaching the single-layer limit. Their chemical properties are no exception, as evidenced by the oxidation and hydrogenation of graphene[1,2]. Given their fully exposed atomic structure, chemically modifying the basal plane of 2D materials provides a particularly promising approach for engineering their properties. However, the chemistry of 2D Transition-Metal Dichalcogenide (TMDC) crystals is mainly defined by their edges[3], where co-ordinatively unsaturated sites prevail. Not coincidentally, these reactive edge sites are also proposed to be responsible for the catalytic activity of $MoS_2$[4,5], one of the most widely studied TMDC materials. However, chemical modification restricted to edges is only efficient in nanoscale islands[6], because the edge-to-surface ratio drastically decreases with increasing lateral dimensions of the sheets. Consequently, for large area 2D crystals it is of particular importance to increase the chemical and catalytic activity of their entire basal plane.

The oxidation of graphene has been studied widely as a promising approach towards its efficient exfoliation[7]. For 2D TMDC crystals the oxidation reaction is also of particular importance, because for some crystals it can spontaneously proceed under ambient conditions. Therefore, its study is essential for understanding their long-term environmental stability, as well for the possibility of creating new routes towards chemically engineering their properties.

While some TMDC crystals, such as $HfSe_2$, $MoTe_2$ or $WTe_2$, are known to be air-sensitive[8,9,10] because they rapidly degrade under ambient conditions, the most widely investigated members of the TMDC family ($MoS_2$, $MoSe_2$, $WS_2$, $WSe_2$) have generally been considered air-stable[11,12], based also on decades-long experience with their bulk crystals. Nevertheless, it has been demonstrated that in single-layer form the oxidation of $MoS_2$ and $WS_2$ also occurs under ambient conditions[13]. This detailed investigation revealed that the oxidation-induced etching observed originates from edges and grain boundaries and proceeds towards the interior of the flakes. Due to the higher strength of the Mo-O bonds, as compared to Mo-S, the substitutional oxidation of the 2D $MoS_2$ basal plane is in principle also thermodynamically favorable[14,15]. However, although such oxidation is a fast, low-barrier process[14] at under-coordinated atomic sites on edges and grain boundaries[15], the oxidation of the defect-free basal plane has been predicted to face relatively high kinetic barriers of about 1.6 eV[16], rendering the basal plane environmentally stable. Although experiments often reveal significantly lower activation-energy values for the $MoS_2$ oxidation, such as 0.54 eV[17] or 0.98 eV[18], no direct evidence for the ambient oxidation of the pristine $MoS_2$ basal plane has been reported so far. We propose that this is mainly due to the inability of the employed structural characterization methods to resolve single-atom level structural changes.

Harsh oxidation processes, such as oxygen plasma treatment, UV-ozone exposure, electrochemical exfoliation, or high temperature (> 300°C) annealing are able to oxidize the basal plane of $MoS_2$ crystals. Based primarily on XPS investigations it was shown that such processes can lead either to covalent oxygen bonding to the top sulfur atoms[14,19], or the formation of completely oxidized $MoO_3$ areas[20,21,22] that can subsequently volatilize, leading to etching. However, neither process is ideal for chemically tuning the properties of $MoS_2$ sheets. The formation of $MoO_3$ destroys the original $MoS_2$ crystal lattice, yielding an overall disordered and fragmented structure, while the chemisorption of oxygen onto chalcogen atoms is predicted to have a relatively weak influence on band structure and properties[14,16]. It has also been shown that oxidation can both enhance and degrade the catalytic activity of $MoS_2$, depending on the structural details[23,24], further emphasizing the importance of investigating and controlling the resulting oxidized structure. The possibility of a substitutional oxidation reaction of $MoS_2$ has also been raised as being thermodynamically preferred to oxygen chemisorption[15,25]; however, no clear experimental evidence has been reported so far. The controlled substitutional oxidation of the $MoS_2$ basal plane would be a highly desirable reaction that preserves the original $MoS_2$ crystal structure, while in contrast to O chemisorption, it is also expected to substantially influence the electronic band structure, enabling a more efficient engineering of its electronic[14,16] and optical[26] properties.

Here we show that the basal plane of the $MoS_2$ monolayers, when subjected to long-term ambient exposure, spontaneously undergo such oxygen-substitution reactions giving rise to a highly crystalline two-dimensional molybdenum oxy-sulfide phase.

## Results and discussion

### Ambient oxidation of the basal plane revealed at single-atom level

We have prepared mechanically exfoliated $MoS_2$ single layers on atomically flat Au (111) substrates using a slightly modified version of a recently developed exfoliation technique[27] yielding single layers with lateral dimension of hundreds of microns (see Supplementary section I. for details). Such large area samples are characterized by an extremely low edge to surface ratio, grain boundary concentration, as well as a much smaller concentration and variety of intrinsic point defects[28], establishing them as an excellent model system for studying the intrinsic chemical properties of the pristine basal plane. The exfoliated $MoS_2$ samples have been stored under ambient conditions (air, room temperature and ambient light) for periods of up to 1.5 years. We have employed STM measurements to follow the atomic level structural changes in the basal plane structure of 2D $MoS_2$ crystals during long-term ambient exposure. So far, mainly optical, scanning electron and atomic force microscopy measurements have been employed to monitor the structural changes induced by oxidation in $MoS_2$ layers, however, the spatial resolution of these methods does not allow the detection of single-atom level modifications. Imaging the atomic-scale structure of the $MoS_2$ basal plane is possible by high resolution scanning Transmission Electron Microscopy (TEM). However, detecting light atoms such as oxygen with TEM is challenging due to their low contrast and easy knock-out[29,30]. By contrast, STM measurements can detect the unaltered structure of such oxygen defects due to its atomic resolution capability and the low energy of the tunneling electrons[31]. Due to the slow oxidation reaction, and the noninvasive nature of the measurements, the structure and number of oxidation induced defect sites did not change during the STM imaging, enabling us to acquire atomic resolution snapshots of the oxidation process.

Atomic resolution STM images of the basal plane of a mechanically exfoliated $MoS_2$ single layer after 1 month and 1 year of ambient exposure are shown in Fig. 1b and 1c, respectively. The STM images have been acquired on the same sample, but at different locations. In contrast to general expectations, STM measurements reveal clear modifications in the atomic structure of the $MoS_2$ basal plane during ambient exposure. Freshly prepared 2D $MoS_2$ crystals contain a native point defect density in the range of $10^{11}$ - $10^{12}$ cm$^{-2}$ (Supplementary Fig. 2). After a month of ambient exposure our STM measurements revealed the formation of new point defects all over the basal plane increasing their concentration into the $3\times10^{12}$ - $2\times10^{13}$ cm$^{-2}$ range (Fig. 1b). The increase of the atomic scale defect concentration (up to $5\times10^{13}$ - $10^{14}$ cm$^{-2}$) appears more strikingly after a year exposure (Fig. 1c), when a substantial area of the sample surface is already covered by such defects, clearly evidencing the ability of the ambient exposure to create new defect sites in the basal plane structure of 2D $MoS_2$ crystals. Although the defect density shows some spatial variation, as evidenced by the concentration ranges provided above, the effect of the ambient exposure time is more significant. Higher resolution STM images (Fig. 1d) shed light on the atomic structure of the defects induced by ambient exposure. Dark triangles in the STM images of 2D $MoS_2$ crystals are characteristic to sulfur atom vacancies [31]. Within the dark triangles, bright spots can be clearly detected that have not been observed before. These central bright spots are a pronounced feature of the experimental STM images, as their contrast is even stronger than that of S atoms of the $MoS_2$ lattice. The most straightforward choice is to attribute the bright spots inside the S vacancies to atoms or molecules saturating the vacancies under ambient conditions. Among the species present under ambient conditions oxygen has been predicted to be the energetically most favorable to saturate an S vacancy [14,16]. To confirm this, we have performed detailed simulations of the STM images of various O-related defects in 2D $MoS_2$ crystals based on DFT calculations of their electronic structure (Supplementary Fig. 3). Our theoretical results reveal that the best agreement with the experimental STM data is provided by single oxygen atoms substituting individual S atoms (saturating the S vacancy) displaying a dark triangle with a bright spot inside. Such a simulated STM image is shown in the inset of Fig.1d. We note that according to both calculations and measurements the contrast of the O substitutions in STM images is dependent on the tip-sample distance; they only become fully apparent at small tip-sample distances (see Supplementary section IV for details). While the agreement of the calculated STM image with experiments is far better for O substitution than for any other calculated defect structure, it is not perfect. This can be attributed partly to the STM image distortions that are unavoidable at room temperature, and partly to the dark 'halo' often emerging near single atomic substitution sites in STM images[32,33]. The latter can be a signature of an electron depletion zone around a negatively charged defect site[34]. STM investigations cannot directly confirm the chemical identity of the incorporated atoms, XPS measurements are usually employed for such purposes. However, due to their low spatial resolution, it was not possible to selectively measure the single layer areas, as exfoliated samples are not homogenously covering the substrate. Furthermore, the relatively low concentration of the incorporated oxygen atoms (< 3 at%) makes such investigations challenging[13]. To gain more information on the nature of the defects progressively forming under ambient conditions, we have also performed Raman spectroscopy (Supplementary Fig. 5) and photoluminescence (PL) measurements. PL spectra on aged samples revealed a new peak at 1.75 eV that is not present in freshly exfoliated samples (see Supplementary Fig. 6). This peak can be clearly associated with vacancy type defect formation[35,36]. Moreover, it has been shown that such peak does not emerge under ultra-high vacuum conditions; the charge transfer induced by the saturation of vacancies with environmental species (oxygen, nitrogen) is necessary[36]. These findings further support our STM and simulation data on the formation of oxygen saturated S vacancies.

The above findings clearly challenge the generally accepted view of environmentally inert $MoS_2$ basal plane, and evidence that the ambient oxidation of the basal plane spontaneously proceeds through formation of S vacancies and their saturation by oxygen. This oxidation process preserves the original crystal lattice with Mo sites in the trigonal prismatic configuration, but coordinated by five sulfur atoms and an oxygen atom. The oxidation speed of the $MoS_2$ basal plane under ambient conditions was found to be of the order of 1 atom / minute / $\mu m^2$. This ultra-slow oxidation reaction in principle enables an extremely precise control of oxygen concentration in the $MoS_2$ lattice. The slow reaction speed is most likely responsible for stabilizing the oxy-sulfide phase against the otherwise favorable full conversion to $MoO_3$[20].

Since the oxidation process is expected to be kinetically limited[14,16], by increasing the temperature, the oxidation speed should also increase. We have subjected a freshly exfoliated $MoS_2$ /Au (111) sample to air inside a furnace heated to 400 K for one week. Subsequent STM investigation revealed a similar defect formation as at room temperature, but the oxygen site concentration reached values of $8 \times 10^{12}$ - $3 \times 10^{13}$ $cm^{-2}$ already after a week of exposure (see Supplementary Fig. 7). This clearly indicates that the oxygen substitution process has been accelerated by increasing the temperature. Furthermore, this observation also excludes the possibility that the ambient oxidation occurs due to the accidental presence of some more reactive oxygen species (e.g., ozone, superoxide, singlet oxygen). Increasing the temperature form 21°C to 127°C is not expected to significantly increase the formation of such reactive species. The accelerated oxygen substitution process at higher temperatures also provides a more feasible route for the synthesis of oxy-sulfide crystals.

The substitutional oxidation of $MoS_2$ basal plane yields a novel 2D $MoS_{2-x}O_x$ solid solution crystal phase (Fig. 1a). The formation of $MoS_{2-x}O_x$ solid solution phase has already been proposed for sputter deposited $MoS_2$ films[37,38,39]. However, so far the structure of the synthesized molybdenum oxy-sulfide films was found to be amorphous or highly disordered with poor long-range crystalline order[40,41]. Here we provide clear evidence for the formation of a highly crystalline 2D molybdenum oxy-sulfide phase and thus a strategy for synthesizing new materials through chemical transformation of 2D crystals that are difficult to synthesize by other methods.

The STM investigations of 2D $MoSe_2$ crystals (prepared and aged under the very same condition as 2D $MoS_2$ crystals) revealed strikingly different behavior upon long-term ambient exposure. Fig. 2 shows atomic resolution STM images of freshly exfoliated and 1 year old $MoSe_2$ single layers. The basal plane of the 2D $MoSe_2$ crystal does not show significant changes even after year-long ambient exposure. Native defects (Se vacancies and chemisorbed adatoms – see the inset of Fig. 2a) are present with a relatively low ($10^{11}$ - $10^{12}$ $cm^{-2}$) density, but their concentration does not increase during ambient exposure.

**The energetics and kinetics of oxygen substitution**

Based on the existing literature, it is difficult to interpret the experimentally observed striking differences in the ambient oxidation behavior of 2D $MoS_2$ and $MoSe_2$ basal planes. However, most theoretical works so far focused on the chemisorption mechanism. Indeed, our experimental results confirm that the pristine basal planes of both $MoS_2$ and $MoSe_2$ single layers are stable against chemisorption or $MoO_3$ transformation. We found that oxygen substitution proceeds in the case of $MoS_2$, but not for $MoSe_2$. In contrast to chemisorption, the oxygen substitution mechanism implies the removal of S (Se) atoms during the oxidation process. To investigate the thermodynamics of this process, we have performed DFT calculations (see Methods for details) regarding the removal of S (Se) atoms from the basal plane of the 2D $MoS_2$ ($MoSe_2$) crystals through oxidation.

First, we have investigated the reaction of surface sulfur atoms of 2D $MoS_2$ crystals with oxygen, yielding volatile $SO_2$ species and creating a sulfur vacancy in the $MoS_2$ basal plane (Fig. 3a). The enthalpy of the reaction is calculated using the formula: $\Delta E$ = E (Reactants) – E (Products) = - 0.49 eV. Since the enthalpy of the final structures is lower than the initial, this implies that S atom removal by oxidation is thermodynamically favorable in the case of $MoS_2$. By contrast, our DFT calculations show that the removal of a Se atom through oxidation of $MoSe_2$ basal plane displays a positive oxidation enthalpy of $\Delta E$ = + 0.75 eV (Fig. 3b). This result evidences the thermodynamically unfavorable nature of selenium oxide formation, which hinders the formation of Se vacancies through oxidation and hence the substitutional oxidation of the $MoSe_2$ basal plane, in agreement with the experimental observations.

Even though thermodynamically favorable, the substitutional oxidation of $MoS_2$ basal plane is still expected to face kinetic barriers. To investigate this, we have performed detailed nudged elastic band (NEB) model calculations[42] for determining the energy of transitional states and finding the potential barriers for the substitutional oxidation process. The results displayed in Fig. 3c show that, in the case of $MoS_2$, the typical kinetic barrier height is about 1 eV for the proposed reaction pathway. We note that by considering the O saturation of the S vacancies the energy of the final state becomes by about - 4 eV lower (Supplementary Fig. 3), indicating the highly favorable nature of the O substitution process as a next step, in accordance with the experimental findings. By contrast, the barriers for the substitutional oxidation of $MoSe_2$ basal plane are about 1.5 eV, with a thermodynamically unfavorable final state (Fig. 3d). Consequently, both energetics and kinetics support our experimental findings on the differences between the oxidation of the 2D $MoS_2$ and $MoSe_2$ basal planes, reflecting the different oxidation mechanisms of sulfur and selenium based compounds. Furthermore, the kinetic barriers of about 1 eV height for the substitutional oxidation of $MoS_2$ are predicted to be surmounted at room temperature on month's timescale, according to transition state theory[43].

Although the basal plane of the $MoSe_2$ appears more stable, the overall environmental stability of the crystals is also dependent on their edge oxidation speed. We found that the oxidation proceeds much faster at the edges of $MoSe_2$ single layers than on the $MoS_2$ edges (Supplementary Figs. 8 and 9). Adsorbed contaminations are naturally present on both $MoS_2$ and $MoSe_2$ surfaces subjected to ambient conditions. However, their role in the ambient oxidation process is expected to be more pronounced at edges and grain boundaries[13,15], as they are known to preferentially attach to these high-energy sites instead of the pristine basal plane.

While structural disorder induced by invasive oxidation is unlikely to be fully reversible, the substitutional oxidation of the $MoS_2$ basal plane does not damage the crystal lattice, so it can in principle be suitable for a fully reversible reduction. Indeed, we found that a simple annealing of the $MoS_{2-x}O_x$ crystals in $H_2S$ atmosphere at 200°C for 30 minutes is able to fully restore the atomic structure of the pure 2D $MoS_2$ phase. Two representative atomic resolution STM images of the $MoS_2$ basal plane before and after the reduction (Fig. 4) clearly evidence that 2D $MoS_{2-x}O_x$ solid solution crystals can be reduced to the pure, almost defect-free $MoS_2$ phase. The feasibility of such a reduction process is also supported by our DFT calculations (Supplementary Fig. 10). The atomically perfect reduction of the oxidized 2D $MoS_{2-x}O_x$ crystals enables a highly efficient engineering of their surface chemistry.

**Catalytic activity of 2D $MoS_{2-x}O_x$ crystals towards hydrogen evolution**

To investigate how the oxygen substitution sites change the properties of $MoS_2$ single layers, we have investigated the catalytic activity of 2D $MoS_{2-x}O_x$ crystals for the electrochemical hydrogen evolution reaction (HER) (see Methods for experimental details). Polarization curve (I-E) measurements have been performed on both $MoS_{2-x}O_x$ and $MoS_2$ single layers. To make the comparison more direct and relevant, we have measured a 1 year old $MoS_{2-x}O_x$ flake before and after its reduction to the pure 2D $MoS_2$ phase. This way, we could measure the catalytic activity of the very same flake both with and without O substitution sites in the basal plane, as revealed by our STM characterization. The measured polarization curves and the corresponding Tafel plots shown in Fig.5 evidence a highly increased catalytic HER activity of the 2D $MoS_{2-x}O_x$ solid solution crystals, as compared to the reduced pure 2D $MoS_2$ phase. Consequently, this enhanced catalytic activity can be clearly related to the presence of substitutional O sites, the only structural feature that displays a correlation with ambient exposure time, according to atomic level structural STM data. We have also investigated the stability of the 2D $MoS_{2-x}O_x$ crystals during 1000 catalytic cycles. After an initial slight decrease by about 10%, the current density stabilizes, indicating a good long-term stability of the catalytic process (Supplementary Fig.11). Confocal Raman microscopy maps revealed that the 2D $MoS_{2-x}O_x$ layer is still continuous after 1000 cycles and the measured Raman spectra remained practically unaltered (Supplementary Fig.12), while atomic-resolution STM measurements confirmed that the O-substitution sites are present after the catalytic process (Supplementary Fig.11).

A widely used parameter for predicting the catalytic activity of various sites is the hydrogen adsorption Gibbs free energy ($\Delta G_H$). We have calculated the $\Delta G_H$ for O substitution sites and compared it to that of pristine $MoS_2$ surface (see Supplementary section X for calculation details). We found that $\Delta G_H$ is lowered to about half on the O sites (+ 1.2 eV) as compared to the S sites (+ 2.2 eV) of the basal plane, which means that hydrogen is much more likely absorbed on the O sites. However, even the substantially reduced $\Delta G_H$ value of about 1 eV is still well above zero, indicating a not very favorable H adsorption. Nevertheless, other effects - not captured by $\Delta G_H$ – can also play an important role in the catalytic activity. Since O atoms have a similar electron configuration to S atoms, it is often assumed that the O substitutions are fully passivating the S vacancies without significantly altering the electronic structure. While O substitutions indeed remove the midgap states characteristic to sulfur vacancies [44], they still substantially change the orbital composition of the valence and conduction bands (see Supplementary section XI), as well as the electron affinity at the substitution sites. The electronegativity and electron affinity of the dopants (including O substitutions) has been shown to play an important role in defining the catalytic activity of graphene [45,46]. To evaluate this effect for the O substitution sites of 2D $MoS_2$ crystals, we have performed the Bader charge analysis (see Supplementary section XII for details) that has proven useful for understanding the catalytic activity of sites where charge transfer plays an important role [47]. The Bader analysis evidenced a strong acceptor-type behavior of the O substitution sites, characterized by almost two-times higher electron affinity (- 0.88 e$^-$) as compared to S atoms (- 0.47 e$^-$). The locally increased electron affinity combined with the experimentally measured overall *n*-doping of the $MoS_2$ crystals (Supplementary Fig. 15) can give rise to localized negative charges on the O substitution sites. These partially screened negative charges can substantially facilitate the adsorption of positively charged H species from the acidic electrolyte. The effect of such charged dopants is not included in the $\Delta G_H$ calculations, as it is challenging to treat charged impurities at the DFT level [48]. Therefore, we propose that the combined effects of the substantially decreased $\Delta G_H$ and increased electronegativity at the O sites — both facilitating the absorption of H species — can give rise to the increased catalytic activity observed in 2D $MoS_{2-x}O_x$ crystals. Understanding the catalytic process is a highly challenging task, largely due to the complexity of the experimentally investigated systems, in strong contrast with the idealized theoretical models[49]. 2D $MoS_{2-x}O_x$ provides an ideal model system for understanding the atomic level relations between active sites and catalytic HER activity, as it is characterized by a single type of active site, with experimentally known atomic structure, while edges and previously investigated more disordered $MoS_2$ structures can host a variety of active sites with complex atomic configurations and often little experimental insight into their precise atomic nature. These findings clearly show that the substitutional oxidation process of the $MoS_2$ basal plane reported here can open new routes for engineering 2D electrocatalysts with single-oxygen-atom active sites of a much higher site density than previously achieved for individual hetero-atom catalysts[50].

## Methods

**Sample preparation.** The investigated $MoS_2$ (and $MoSe_2$) single layers have been prepared by mechanical exfoliation of bulk $MoS_2$ ($MoSe_2$) synthetic crystals of high structural quality (2DSemiconductors). We have employed a slightly modified version (see Supplementary section I) of the mechanical exfoliation technique developed by us and discussed in details in ref. *(27)*, providing single layer TMDC flakes with hundreds of microns lateral dimensions on atomically flat Au (111) surfaces. The single layer nature of the investigated $MoS_2$ and $MoSe_2$ crystals has also been confirmed by Raman spectroscopy. The samples have been stored under ambient laboratory conditions: in air, at room temperature and ambient light conditions.

**Characterization.** Atomic resolution STM measurements have been performed on a Nanoscope E STM operating under ambient conditions. Tunneling spectroscopy data have been performed in an RHK PanScan STM under UHV conditions at room temperature. Large area flakes could be easily identified under an optical microscope enabling the guided landing of the STM tip on MoS$_2$ single layers. Atomic resolution imaging could be achieved under ambient conditions with typical parameter ranges of |U$_{bias}$| = 5 - 50 mV, and I$_{tunnel}$ = 1 - 3 nA. Although the resolution of the hexagonal lattice of the top layer of S atoms can be routinely achieved, atomic resolution of individual point defects is challenging. Defects were clearly resolved when imaging at bias voltages (energies) within the band gap. This is possible due to the influence of the Au(111) substrate, which induce a small but finite density of states within the band gap (confirmed by our tunneling spectroscopy measurements), conferring a weak metallic character to the supported MoS$_2$ single layers. Raman measurements have been conducted on a confocal Raman microscope (Witec 300RSA) with $\lambda$ = 532 nm and W = 1mW. We have ensured that during the Raman measurement the spectra of the sample did not change detectably.

**Electrochemical measurements.** The measurements have been conducted on selected sample areas (0.4 – 0.8 mm diameter) of a single 2D MoS$_2$ flake supported by a 100 nm thick Au(111) film on a glass substrate. Control measurements on the same Au (111) substrate as well as a Pt plate have been conducted in the same experimental configuration. To compare the performance of different samples in HER, linear sweep voltammetry was performed in a three-electrode configuration using 0.5M sulfuric acid electrolyte at room temperature. Silver / silver chloride and Pt wire were used as counter and reference electrodes, respectively. Potential sweeps were acquired at a scan rate of 2 mV/s using a Bio-Logic SP-150 potentiostat.

**Computational details.** All calculations were performed on TMDC single layers in the framework of spin-polarized DFT theory implemented in the VASP software package, using the plane wave basis set and projector augmented wave method. Exchange-correlation effects were taken into account in the framework of local density approximation (LDA) and generalized gradient approximation (GGA) by Perdew–Burke–Ernzerhof (PBE) functional.

The defects were modelled in periodically repeated 8×8 supercells. The Brillouin zone of the supercells was sampled with (2×2×1) Monkhorst−Pack mesh of k-points for geometry optimization and (4×4×1) for STM-image calculations. The cutoff energy for the plane-wave basis set was set to be 400 eV. To avoid artificial interactions between periodic replicas of low-dimensional nanoclusters, a vacuum interval of 15 Å was introduced in all supercell geometry. The nudged elastic band (NEB) method was applied for modeling transitional states and finding potential barriers of $MoS_2$ and $MoSe_2$ oxidation.

**Competing interests**

The authors declare no competing financial and non-financial interests.

**Data availability**

The data supporting the findings of this study are available within the article and its Supplementary Information files. All other relevant source data are available from the corresponding author upon request.

**Acknowledgements**

The work has been performed in the framework of the NanoFab2D ERC Starting grant, the H2020 Graphene Core2 project no. 785219 and the Korea Hungary Joint Laboratory for Nanosciences. L.T. acknowledges OTKA grant K108753 and the "Lendület" Program. The work was also supported by the VEKOP-2.3.2-16-2016-00011grant supported by the European Structural and Investment Funds. Z.I.P. and P.B.S. gratefully acknowledge the financial support of the Ministry of Education and Science of the Russian Federation in the framework of Increase Competitiveness Program of NUST "MiSIS" (No. K2-2017-001). Z.I.P. and P.B.S. are grateful to the supercomputer cluster "NUST-MiSIS" provided by the Materials Modelling and Development Laboratory (supported via the Grant from the Ministry of Education and Science of the Russian Federation No. 14.Y26.31.0005), to the Information Technology Centre of Novosibirsk State University for providing access to the cluster computational resources. P.V. acknowledges the "Plateforme Technologique de Calcul Intensif (PTCI)," which was supported by the F.R.S.-FNRS under Convention No. 2.5020.11. P.B.S. acknowledges the financial support of the RFBR, according to the research project No. 16-32-60138 mol_a_dk. We are grateful for J.S. Pap for useful discussions on electrochemistry.

**Author contributions**. L.T. conceived and designed the experiments. J.P. and G.Z.M. prepared the samples and performed the STM measurements. T.O. performed the chemical reduction and electrocatalytic experiments. Z.I.P, P.V., and P.B.S. performed the theoretical calculations. G.D. and G.Z.M. conducted the Raman and PL investigations. L.T., P.B.S. and C.H. supervised the project. L.T. wrote the paper. All authors discussed the results and commented on the manuscript.

**Correspondence and requests for materials** should be addressed to L.T.

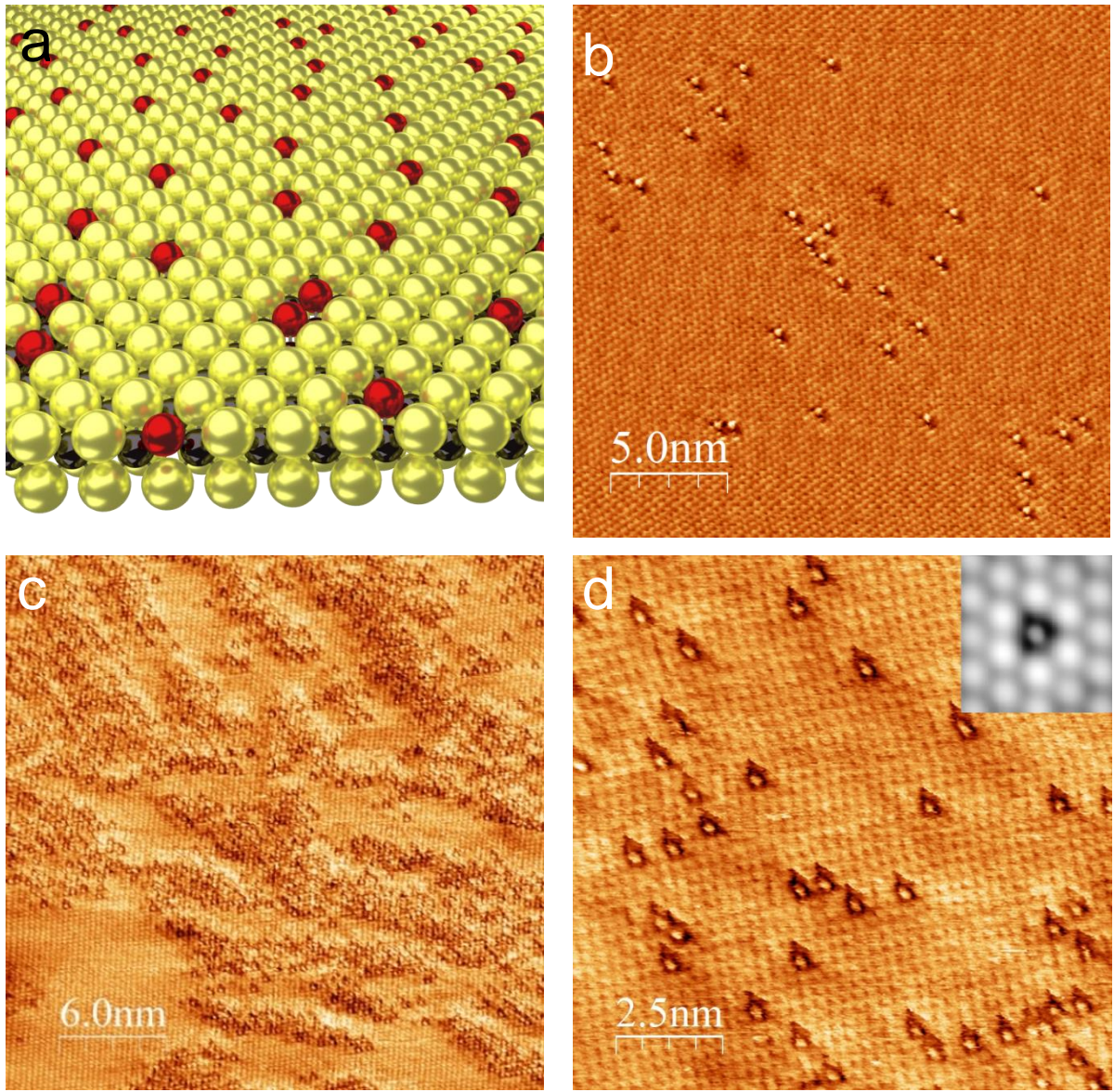

**Figure 1**. **2D MoS$_{2-x}$O$_x$ solid solution crystals by ambient oxidation of MoS$_2$**. *a) Schematic atomic structure of the MoS$_{2-x}$O$_x$ solid solution monolayer. b)-d) Atomic resolution STM images (5mV, 2nA) of an exfoliated MoS$_2$ single layer after one month (b) and one year (c) ambient exposure revealing a progressive defect formation. d) Higher resolution STM image displaying the incorporation of the oxygen atoms (bright spots) into the S vacancies (dark triangles). Inset shows simulated STM image based on DFT calculations of an oxygen saturated S vacancy site in the 2D MoS$_2$ crystal.*

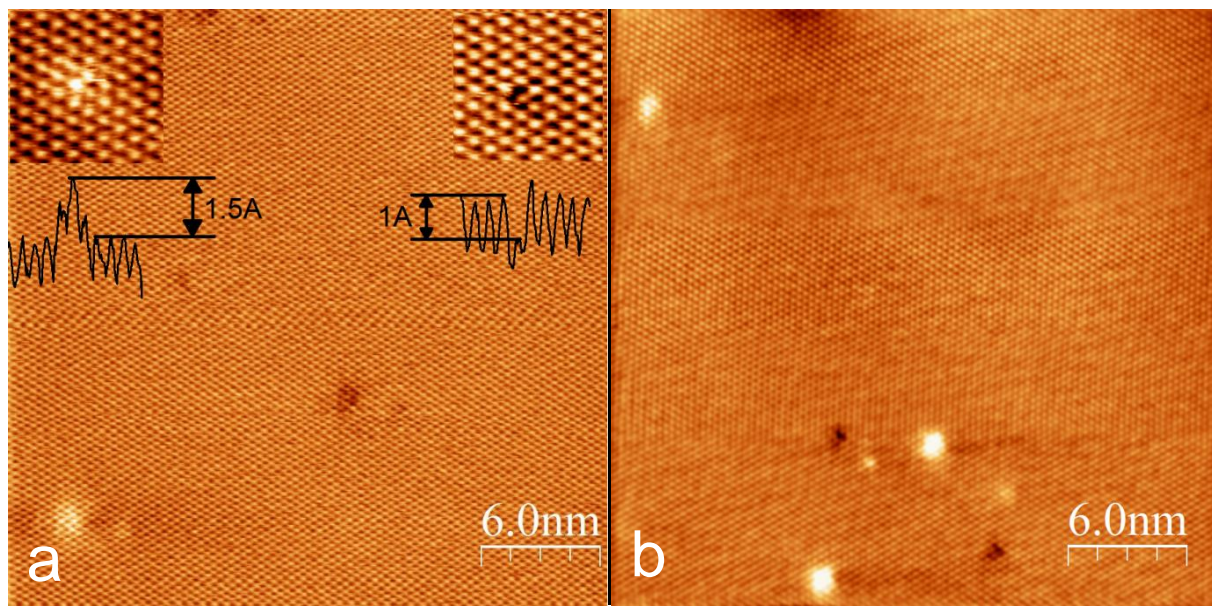

**Figure 2. Stability of 2D MoSe₂ basal plane under ambient conditions**. *Atomic resolution STM images (5mV, 1nA) of 2D MoSe₂ crystals on Au (111) substrate freshly prepared (a) and after 1 year of ambient exposure (b), revealing a remarkable stability of the basal plane under ambient conditions. The insets of panel (a) show higher resolution STM images and the corresponding line cuts of bright (chemisorbed adatom) and dark (Se vacancy) defect sites. The number of these defects does not increase with ambient exposure time.*

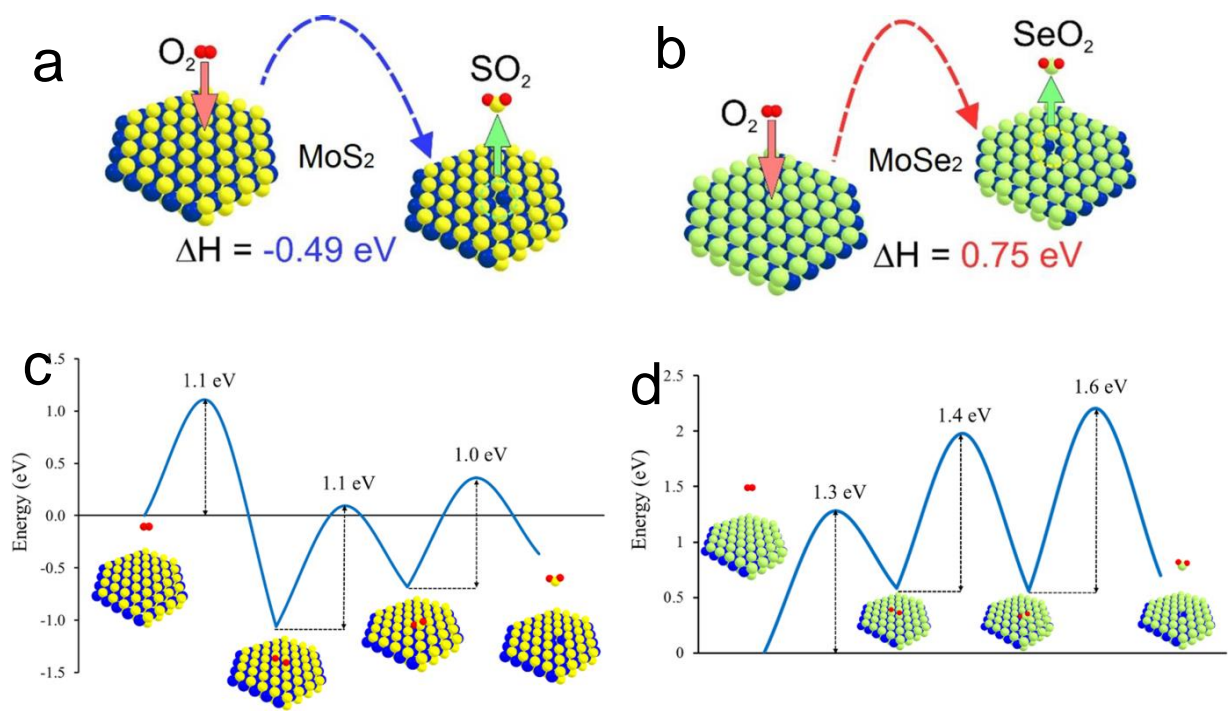

**Figure 3. Energetics and kinetics of the oxygen substitution in the MoS$_2$ and MoSe$_2$ basal plane.** *The process of chalcogenide atom vacancy formation through oxidation is characterized by a negative oxidation enthalpy for the defect free 2D MoS$_2$ basal plane (a) and a positive enthalpy for MoSe$_2$, protected from oxidation by the endothermic nature of SeO$_2$ formation (b). Kinetic energy barriers calculated by the nudged elastic band model for 2D MoS$_2$ (c) and 2D MoSe$_2$ (d) crystals reveal substantially lower barriers for MoS$_2$ of about 1 eV height that can be overcome even at room temperature on months-long time scale.*

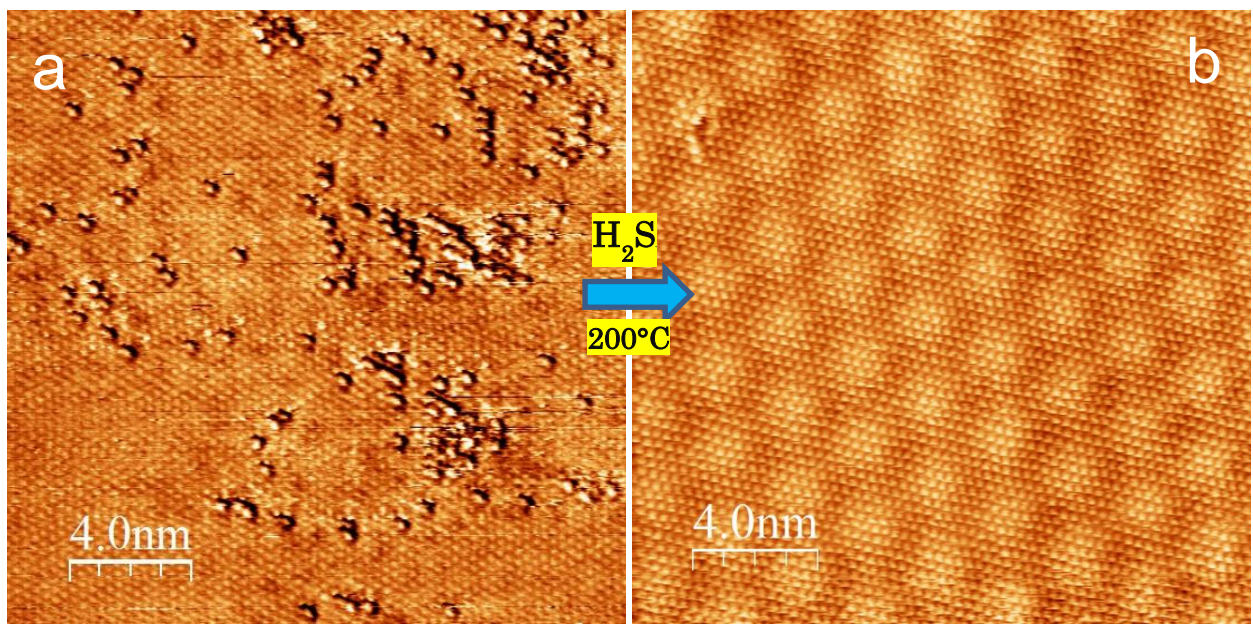

**Figure 4**. **Reduction of 2D MoS$_{2-x}$O$_x$ to pristine MoS$_2$.** *Representative atomic resolution STM images (5mV, 2nA) of 2D MoS$_{2-x}$O$_x$ before (a) and after (b) 30 minutes annealing at 200° C in H$_2$S atmosphere, evidencing the atomically perfect reduction of the oxy-sulfide solid solution to the pure MoS$_2$ phase through the re-substitution of the single atomic O sites by S atoms.*

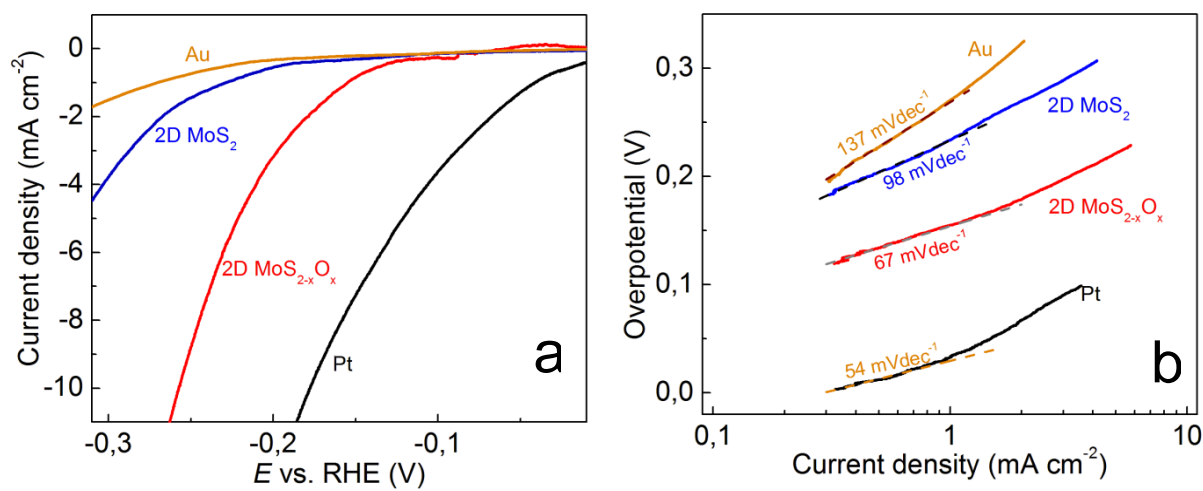

**Figure 5**. **Catalytic activity of 2D MoS$_{2-x}$O$_x$ for hydrogen evolution.** *Linear sweep votammogram curves (a) and the corresponding Tafel plots (b) for: Au substrate, MoS$_2$ single layer, MoS$_{2-x}$O$_x$ single layer (1-year-old), and Pt substrate, revealing a significantly higher catalytic activity of the 2D oxy-sulfide phase as compared to the pure MoS$_2$ phase. The increased catalytic activity can be attributed to single-atomic O sites progressively incorporating into the MoS$_2$ basal plane during ambient exposure.*